# A New Lever Function with Adequate Indeterminacy [*]

Shenghui Su [1, 5], Ping Luo [2], Shuwang Lü [3, 5], and Maozhi Xu [4]

[1] College of Computers, Nanjing Univ. of Aeronautics & Astronautics, Nanjing 211106, PRC
[2] Software School, Tsinghua University, Beijing 100084, PRC
[3] Laboratory of Information Security, Chinese Academy of Sciences, Beijing 100039, PRC
[4] School of Mathematics, Peking University, Beijing 100871, PRC
[5] Laboratory of Computational Complexity, BFID Corporation, Fuzhou 350207, PRC

**Abstract**: The key transform of the REESSE1+ asymmetrical cryptosystem is $C_i \equiv (A_i W^{\ell(i)})^\delta$ (% $M$) with $\ell(i) \in \Omega = \{5, 7, \ldots, 2n+3\}$ for $i = 1, \ldots, n$, where $\ell(i)$ is called a lever function. In this paper, we give a simplified key transform $C_i \equiv A_i W^{\ell(i)}$ (% $M$) with a new lever function $\ell(i)$ from $\{1, \ldots, n\}$ to $\Omega_\pm = \{\pm 5, \pm 6, \ldots, \pm(n+4)\}$. Discuss the necessity of the new $\ell(i)$, namely that a simplified private key is insecure if the new $\ell(i)$ is a constant but not one-to-one function. Further, expound the sufficiency of the new $\ell(i)$ from four aspects: (1) indeterminacy of the new $\ell(i)$, (2) insufficient conditions for neutralizing the powers of $W$ and $W^{-1}$ even if $\Omega_\pm = \{5, \ldots, n+4\}$, (3) verification by examples, and (4) the running times of continued fraction attack and W-parameter intersection attack which are the two most efficient algorithms of probabilistic polytime attacks so far. Last, we detail the relation between a lever function and a random oracle.

**Keywords**: Asymmetrical cryptosystem, Coprime sequence, Lever function, Continued fraction attack, Random oracle

## 1    Introduction

Theories of computational complexity such as the class P, class NP, one-way functions, and trapdoor functions provide asymmetrical cryptosystems with foundation stones [1][2][3]. For instance, the RSA cryptosystem is founded on the integer factorization problem (IFP) [4], and ElGamal is founded on the discrete logarithm problem (DLP) [5]. It appeals to people whether polytime algorithms for solving IFP and DLP on electronic computers exist or not since IFP and DLP are not proved NP-complete.

To $N = pq$ with prime $p$ and $q$, if $N$ is given, the values of $p$ and $q$ are determined. To $y \equiv g^x$ (% $p$) with $g$ a generator of $(\mathbb{Z}_p^*, \cdot)$, if $y$ is given, the value of $x$ is also determined. On the other hand, there exists such a class of computational problems that they looks very ordinary but brings indeterminacy into asymmetrical cryptosystems — a permutation problem for example.

In the REESSE1+ cryptosystem [6], the key transform is $C_i \equiv (A_i W^{\ell(i)})^\delta$ (% $M$) with $\ell(i) \in \{5, 7, \ldots, 2n+3\}$. A private key $\langle\{A_i\}, \{\ell(i)\}, W, \delta\rangle$ is undoubtedly secure due to the existence of the random $\delta$ ($\in [1, M-1]$)[6]. If let $\delta = 1$ (namely $C_i \equiv A_i W^{\ell(i)}$ (% $M$)) and $\ell(i) \in \{\pm 5, \pm 6, \ldots, \pm(n+4)\}$, what about it?

In this paper, we will investigate the effect of a new lever function $\ell(.)$ from $\{1, \ldots, n\}$ to $\{\pm 5, \ldots, \pm(n+4)\}$ on the security of a simplified transform $C_i \equiv A_i W^{\ell(i)}$ (% $M$), where $M$ is a prime.

Throughout the paper, unless otherwise specified, $n$ ($\geq 80$) is the bit-length of a plaintext block, the sign % denotes "modulo", $\bar{M}$ does ($M-1$), $\pm$ means the selection of a plus or minus sign, $\lg x$ does a logarithm of $x$ to the base 2, $\neg x$ does the opposite of a bit $x$, $\mathcal{P}$ does the maximal prime allowed in coprime sequences, $|x|$ does the absolute value of an integer $x$, $|S|$ does the size of a set $S$, and $\gcd(a, b)$ represents the greatest common divisor. Without ambiguity, "% $M$" is usually omitted in expressions.

## 2    Simplified REESSE1+ Encryption Scheme

To inspect the indeterminacy of the new $\ell(.)$ from $\{1, \ldots, n\}$ to $\{\pm 5, \ldots, \pm(n+4)\}$, let $\delta = 1$ in the key transform of REESSE1+, and thus we acquire the simplified REESSE1+ encryption scheme.

### 2.1   Two Definitions

***Definition 1:*** If $A_1, A_2, \ldots, A_n$ are $n$ pairwise distinct positive integers such that $\forall A_i, A_j$ ($i \neq j$), either $\gcd(A_i, A_j) = 1$, or $\gcd(A_i, A_j) = F \neq 1$ with $(A_i / F) \nmid A_k$ and $(A_j / F) \nmid A_k$ $\forall k$ ($\neq i, j$) $\in [1, n]$, then this integer progression is called a coprime sequence, denoted by $\{A_1, ..., A_n\}$, and shortly $\{A_i\}$.

---

[*] This work is supported by MOST with Project 2009AA01Z441. Corresponding email: junaplus@hotmail.com.

Notice that the elements of a coprime sequence are not necessarily pairwise coprime, but a sequence whose elements are pairwise coprime must be a coprime sequence.

**Property 1:** Let $\{A_1, ..., A_n\}$ be a coprime sequence. If randomly select $m \in [1, n]$ elements $A_{x_1}, ..., A_{x_m}$ from the sequence, then the mapping from a subset $\{A_{x_1}, ..., A_{x_m}\}$ to a subset product $G = \prod_{i=1}^{m} A_{x_i}$ is one-to-one, namely the mapping from $b_1...b_n$ to $G = \prod_{i=1}^{n} A_i^{b_i}$ is one-to-one, where $b_1...b_n$ is a bit string.

Refer to [6] for its proof.

**Definition 2:** The secret parameter $\ell(i)$ in the key transform of an asymmetrical cryptosystem is called a lever function, if it has the following features:

1) $\ell(.)$ is an injection from the domain $\{1, ..., n\}$ to the codomain $\Omega \subset \{5, ..., \bar{M}\}$, where $\bar{M}$ is large;
2) the mapping between $i$ and $\ell(i)$ is established randomly without an analytical expression;
3) an attacker has to be faced with all the arrangements of $n$ elements in $\Omega$ when extracting a related private key from a public key;
4) the owner of a related private key only needs to consider the accumulative sum of $n$ elements in $\Omega$ when recovering a related plaintext from a ciphertext.

The Feature 3 and 4 manifest that if $n$ is large enough, it is infeasible for the attacker exhaustively to search all the permutations of elements in $\Omega$ while the decryption of a normal ciphertext is feasible in some polytime. Thus, there are the large amount of calculation on $\ell(.)$ at "a public terminal", and the small amount of calculation on $\ell(.)$ at "a private terminal".

Notice that ① the number of elements of $\Omega$ is not less than $n$; ② considering the speed of decryption, the absolute values of all the elements should be relatively small; ③ the lower limit 5 will make seeking the root $W$ from $W^{\ell(i)} \equiv A_i^{-1} C_i \ (\% \ M)$ face unsolvability when the value of $A_i \ (\leq 1201)$ is guessed [7].

## 2.2 Key Generation Algorithm

In the simplified REESSE1+ scheme, substitute $\Omega = \{5, 7, ..., 2n+3\}$ with $\Omega_\pm = \{\pm 5, \pm 6, ..., \pm(n+4)\}$.
Let $|\Omega_\pm|$ be the set of absolute values of all the elements in $\Omega_\pm$.
Let $\Lambda = \{2, ..., \mathcal{P}\}$, where $\mathcal{P} = 863, 937, 991,$ or $1201$ when $n = 80, 96, 112,$ or $128$.
This algorithm is employed by a certificate authority or the owner of a key pair.
INPUT: the integer $n$; the set $\Lambda$.
S1: Give $\Omega_\pm$ a random permutation of $\{\pm 5, ..., \pm(n + 4)\}$.
S2: Randomly produce odd and coprime $A_1, ..., A_n \in \Lambda$.
S3: Find a prime $M > \prod_{i=1}^{n} A_i$ making $q^2 \mid \bar{M} \ \forall \ q$ (a prime) $\in |\Omega_\pm|$.
S4: Stochastically pick an integer $W \in (1, \bar{M})$.
S5: Stochastically yield pairwise distinct $\ell(1), ..., \ell(n) \in \Omega_\pm$.
S6: Compute $C_i \leftarrow A_i W^{\ell(i)} \ \% \ M$ for $i = 1, ..., n$.
OUTPUT: a public key $\langle \{C_1, ..., C_n\}, M \rangle$; a private key $\langle \{A_1, ..., A_n\}, W, M \rangle$.
The secret parameter $\{\ell(1), ..., \ell(n)\}$ may be discarded.
Notice that in arithmetic modulo $\bar{M}$, the negative integer $-x$ represents $\bar{M} - x$.

## 2.3 Encryption Algorithm

This algorithm is employed by a person who wants to encrypt plaintexts.
INPUT: a public key $\langle \{C_1, ..., C_n\}, M \rangle$; an $n$-bit plaintext block $b_1...b_n$.
S1: Set $\bar{G} \leftarrow 1, i \leftarrow 1$.
S2: If $b_i = 1$ then let $\bar{G} \leftarrow \bar{G} C_i \ \% \ M$.
S3: Let $i \leftarrow i + 1$.
S4: If $i \leq n$ then goto S2; else end.
OUTPUT: the ciphertext $\bar{G} \equiv \prod_{i=1}^{n} C_i^{b_i} \ (\% \ M)$.

**Definition 3:** Given $\bar{G}$ and $\langle \{C_1, ..., C_n\}, M \rangle$, seeking $b_1...b_n$ from $\bar{G} \equiv \prod_{i=1}^{n} C_i^{b_i} \ (\% \ M)$ is called a subset product problem, shortly SPP [6][8].

Notice that when $\lceil \lg \bar{M} \rceil < 1024$, a discrete logarithm can be found in tolerable subexponential time.

Let $g$ be a generator of $(\mathbb{Z}_M^*, \cdot)$, $\bar{G} \equiv g^u \ (\% \ M)$, $C_1 \equiv g^{v_1} \ (\% \ M), ..., C_n \equiv g^{v_n} \ (\% \ M)$, and then a SPP $\bar{G} \equiv \prod_{i=1}^{n} C_i^{b_i} \ (\% \ M)$ is degenerated to a subset sum problem $u \equiv b_1 v_1 + ... + b_n v_n \ (\% \ \bar{M})$.

Because the knapsack density from this subset sum problem is less than 1, a simplified REESSE1+ ciphertext $\bar{G}$ is not robust [9], which indicates that only if $\lceil \lg \bar{M} \rceil \geq 1024$, can the simplified REESSE1+ cryptosystem have practical sense.

### 2.4 Decryption Algorithm

This algorithm is employed by a person who wants to decrypt ciphertexts.
INPUT: a private key $\langle \{A_1, ..., A_n\}, W, M \rangle$; a ciphertext $\bar{G}$.
S1: Set $X_0 \leftarrow \bar{G}, X_1 \leftarrow \bar{G}, h \leftarrow 0$.
S2: If $2 \mid X_h$ then $X_h \leftarrow X_h W^{(-1)^h} \% M$, goto S2; else next.
S3: Set $b_1...b_n \leftarrow 0, G \leftarrow X_h, i \leftarrow 1$.
S4: If $A_i \mid G$ then let $b_i \leftarrow 1, G \leftarrow G / A_i$.
S5: Let $i \leftarrow i + 1$. If $i \leq n$ and $G \neq 1$ then goto S4.
S6: If $G \neq 1$ then do $h \leftarrow \neg h, X_h \leftarrow X_h W^{(-1)^h} \% M$, goto S2; else end.
OUTPUT: a plaintext block $b_1...b_n$.
Notice that only if $\bar{G}$ is a true ciphertext, can this algorithm terminates normally.

## 3 Necessity of the Lever Function $\ell(.)$

We will discuss that the new lever function $\ell(.)$ from $\{1, ..., n\}$ to $\Omega_\pm = \{\pm 5, ..., \pm(n+4)\}$ is necessary for $C_i \equiv A_i W^{\ell(i)} (\% M)$ to resist continued fraction attack and W-parameter intersection attack.

The necessity of the new $\ell(.)$ means that if a simplified REESSE1+ private key is secure, $\ell(.)$ as a one-to-one function must exist in the key transform. The equivalent contrapositive assertion is that if $\ell(.)$ as a one-to-one function does not exist (but $\ell(i)$ as a constant function exists) in the key transform, a simplified private key will be insecure.

### 3.1 Continued Fraction Attack on a Simplified Private Key

**Theorem 1:** If $\alpha$ is an irrational number, $r, s > 0$ are two integers, and $r/s$ is a rational in the lowest terms such that $|\alpha - r/s| < 1/(2s^2)$, then $r/s$ is a convergent of the simple continued fraction expansion of $\alpha$.

Refer to [10] for the proof.

Notice that theorem 1 also holds when $\alpha$ is a rational number [10].

For an asymmetrical cryptosystem, if a private key is insecure, a plaintext must be insecure. Hence, the security of a private key is most foundational [11].

**Definition 4:** Attack on $C_i \equiv A_i W^{\ell(i)} (\% M)$ with $\ell(i) \in \Omega_\pm = \{\pm 5, ..., \pm(n+4)\}$ for $i = 1, ..., n$ by a convergent of the continued fraction of $G_z / M$, where $G_z \equiv (C_{x_1}...C_{x_m})(C_{y_1}...C_{y_h})^{-1}$ with $m \in [1, n-1]$, $h \in [1, n-m]$, and $x_j \neq y_k \forall j \in [1, m]$ and $k \in [1, h]$, is called continued fraction attack.

**Property 2:** Let $\bar{e} \in [1, \bar{M}]$ be any constant integer. If the key transform of the simplified REESSE1+ cryptosystem is $C_i \equiv A_i W^{\bar{e}} (\% M)$, namely $\ell(i) = \bar{e}$ for $i = 1, ..., n$, a simplified REESSE1+ private key $(\{A_1, ..., A_n\}, W^{\bar{e}})$ is insecure.

*Proof.*

Assume that $\ell(1) = \ell(2) = ... = \ell(n) = \bar{e}$, where $\bar{e}$ is a constant integer.

Then, the key transform becomes as
$$C_i \equiv A_i W^{\bar{e}} (\% M),$$
and especially when $\bar{e} = 1$, $C_i \equiv A_i W (\% M)$ for $i = 1, ..., n$.

Since $(\mathbb{Z}_M^*, \cdot)$ is an Abelian group [7], of course, there is
$$C_i^{-1} \equiv (A_i W^{\bar{e}})^{-1} (\% M).$$

$\forall x \in [1, n-1]$, let
$$G_z \equiv C_x C_n^{-1} (\% M).$$

Substituting $A_x W^{\bar{e}}$ and $A_n W^{\bar{e}}$ respectively for $C_x$ and $C_n$ in the above congruence yields
$$G_z \equiv A_x W^{\bar{e}} (A_n W^{\bar{e}})^{-1} (\% M)$$
$$A_n G_z \equiv A_x (\% M)$$
$$A_n G_z - LM = A_x,$$
where $L$ is a positive integer.

The either side of the equation is divided by $A_n M$ gives
$$G_z / M - L / A_n = A_x / (A_n M). \qquad (1)$$

Due to $M > \prod_{i=1}^{n} A_i$ and $A_i \geq 2$, there is
$$G_z / M - L / A_n < A_x / (A_n \prod_{i=1}^{n} A_i)$$
$$= A_x / (A_n^2 \prod_{i=1}^{n-1} A_i) \leq 1 / (2^{n-2} A_n^2),$$

that is,
$$G_z/M - L/A_n < 1/(2^{n-2} A_n^2). \tag{2}$$
Evidently, as $n > 2$, there is
$$G_z/M - L/A_n < 1/(2 A_n^2). \tag{2'}$$
In terms of theorem 1, $L/A_n$ is a convergent of the continued fraction of $G_z/M$.

Thus, $L/A_n$, namely $A_n$ may be determined by (2′) in polytime since the length of the continued fraction will not exceed $\lceil \lg M \rceil$, and further $W^{\bar{e}} \equiv C_n A_n^{-1}$ (% $M$) may be computed, which indicates the original coprime sequence $\{A_1, …, A_n\}$ with $A_i \leq \mathcal{P}$ can almost be recovered. □

The $W$ in every $C_i$ has the same exponent, and in addition the powers of $W$ and $W^{-1}$ in any $C_x C_n^{-1}$ % $M$ always counteract each other, so there does not exist the indeterministic reasoning method when $\ell(i)$ is the constant integer $\bar{e}$.

It should be noted that when a convergent of the continued fraction of $G_z/M$ satisfies (2′), the some subsequent convergents also possibly satisfies (2′), and if so, it will bring about the nonuniqueness of value of $A_n$. Therefore, we say that $\{A_1, …, A_n\}$ with $A_i \leq \mathcal{P}$ can almost be recovered.

### 3.2  W-parameter intersection Attack on a Simplified Private Key

Assume that $\ell(1) = … = \ell(n) = \bar{e}$, where $\bar{e}$ is a constant integer. Then the key transform is $C_i \equiv A_i W^{\bar{e}}$ (% $M$) for $i = 1, …, n$. Hence, there exists the following attack which is described with an algorithm.

INPUT: a public key $\langle \{C_1, …, C_n\}, M \rangle$
S1: For $i = 1, …, n$ do
    while $A_i$ traverses $\Lambda$ do
        S1.1: compute $W^{\bar{e}}$ such that $W^{\bar{e}} \equiv C_i A_i^{-1}$ (% $M$);
        S1.2: place a tuple $\langle W^{\bar{e}}, A_i \rangle$ into the set $\bar{V}_i$.
S2: Seek the intersection $\bar{V} = \bar{V}_1 \cap … \cap \bar{V}_n$ on $W^{\bar{e}}$. (Note $1 \leq |\bar{V}| < |\Lambda|$)
S3: Extract $W^{\bar{e}}$ from $\bar{V}$ and corresponding $A_i$ from $\bar{V}_i$.
S4: If $A_1, …, A_n$ are pairwise coprime then $W^{\bar{e}}$ and $\{A_i\}$ valid.
OUTPUT: a private key $\langle \{A_1, …, A_n\}, W^{\bar{e}} \rangle$.

It is not difficult to understand that the time complexity of the above attack is dominantly involved in S1 and S2. Concretely speaking, the time complexity is $O(2|\Lambda|n)$, and polynomial in $n$.

Section 3.1 and 3.2 manifest that when every $\ell(i)$ is a constant integer $\bar{e}$, a related private key can be deduced from a public key, and further a related plaintext can be inferred from a ciphertext. Thus, the one-to-one lever function $\ell(.)$ is necessary to the security of a simplified REESSE1+ private key.

## 4  Sufficiency of the Lever Function $\ell(.)$

The sufficiency of the new lever function $\ell(.)$ ($\{1,…,n\} \to \Omega_\pm = \{\pm 5,…,\pm(n+4)\}$) for $C_i \equiv A_i W^{\ell(i)}$ (% $M$) to resist continued fraction attack and W-parameter intersection attack, which are the two most efficient of the probabilistic polytime attack algorithms so far, means that if $\ell(1), …, \ell(n) \in \Omega_\pm$ are pairwise distinct, a simplified REESSE1+ private key will be secure.

We will see that continued fraction attack and W-parameter intersection attack are ineffectual on the security of a private key when $\Omega_\pm$ is indeterminate, and even if $\Omega_\pm = \{5, …, n + 4\}$ is taken by machines and known to adversaries, continued fraction attack does not always threaten $C_i \equiv A_i W^{\ell(i)}$ (% $M$).

### 4.1  Indeterminacy of the Lever Function $\ell(.)$

According to Section 2.2, if the lever function $\ell(.)$ exists, we have
$$C_i \equiv A_i W^{\ell(i)} \ (\% \ M),$$
where $A_i \in \Lambda = \{2, …, \mathcal{P}\}$, and $\ell(i) \in \Omega_\pm = \{\pm 5, …, \pm(n + 4)\}$ for $i = 1, …, n$.

The lever function $\ell(.)$ brings adversaries at least two difficulties:
- No method by which one can directly judge whether the power of $W$ in $C_{x_1}…C_{x_m}$ counteracts the power of $W^{-1}$ in $(C_{y_1}…C_{y_h})^{-1}$ or not;
- No criterion by which one can verify an indeterministic reasoning presupposition in polytime.

The indeterministic reasoning based on continued fractions means that ones first presuppose that the powers of the parameter $W$ and the inverse $W^{-1}$ counteract each other in a product, and then judge whether the presupposition holds or not by the consequence.

According to Section 3, first select $m$ ($\in [1, n - 1]$) elements and $h$ ($\in [1, n - m]$) other elements from $\{C_1, \ldots, C_n\}$. Let
$$G_x \equiv C_{x_1} \ldots C_{x_m} \ (\% \ M),$$
$$G_y \equiv C_{y_1} \ldots C_{y_h} \ (\% \ M),$$
where $x_j \neq y_k \ \forall j \in [1, m]$ and $k \in [1, h]$. Let
$$G_z \equiv G_x G_y^{-1} \ (\% \ M).$$

Since $\{\ell(1), \ldots, \ell(n)\}$ is any arrangement of $n$ elements in $\Omega_\pm$, it is impossible to predicate that $G_z$ does not contain the factor $W$ or $W^{-1}$. For a further deduction, we have to *presuppose* that the power of $W$ in $G_x$ is exactly counteracted by the power of $W^{-1}$ in $G_y^{-1}$, and then,
$$G_z \equiv (A_{x_1} \ldots A_{x_m})(A_{y_1} \ldots A_{y_h})^{-1} \ (\% \ M)$$
$$G_z (A_{y_1} \ldots A_{y_h}) \equiv A_{x_1} \ldots A_{x_m} \ (\% \ M)$$
$$G_z (A_{y_1} \ldots A_{y_h}) - LM = A_{x_1} \ldots A_{x_m}$$
$$G_z / M - L / (A_{y_1} \ldots A_{y_h}) = (A_{x_1} \ldots A_{x_m}) / (M A_{y_1} \ldots A_{y_h}),$$
where $L$ is a positive integer.

Denoting the product $A_{y_1} \ldots A_{y_h}$ by $\bar{A}_y$ yields
$$G_z / M - L / \bar{A}_y = (A_{x_1} \ldots A_{x_m}) / (M \bar{A}_y). \tag{3}$$
Due to $M > \prod_{i=1}^n A_i$ and $A_i \geq 2$, we have
$$G_z / M - L / \bar{A}_y < 1 / (2^{n-m-h} \bar{A}_y^2). \tag{4}$$
Obviously, when $n > m + h$, (4) may have a variant, namely
$$G_z / M - L / \bar{A}_y < 1 / (2 \bar{A}_y^2). \tag{4'}$$
Notice that when $n = m + h$, if $M > 2(\prod_{i=1}^n A_i)$, (4') still holds.

Especially, when $n > 3$, $h = 1$, and $m = 2$, there exists
$$G_z / M - L / A_{y_1} < 1 / (2^{n-3} A_{y_1}^2) < 1 / (2 A_{y_1}^2). \tag{4''}$$
Obviously, as a discriminant, (4) is stricter than (4') and (4''). (4'') is consistent with theorem 1.

**Property 3**: Let $h + m \leq n$. If $\ell(x_1) + \ldots + \ell(x_m) = \ell(y_1) + \ldots + \ell(y_h)$, the subset product $\bar{A}_y = A_{y_1} \ldots A_{y_h}$ in (4') will be found in polytime.

*Proof.*

$\ell(x_1) + \ldots + \ell(x_m) = \ell(y_1) + \ldots + \ell(y_h)$ means that the exponent on $W$ in $C_{x_1} \ldots C_{x_m}$ is counteracted by the exponent on $W^{-1}$ in $(C_{y_1} \ldots C_{y_h})^{-1}$, and thus (4') holds.

In terms of theorem 1, $L / \bar{A}_y$ is inevitably a convergent of the continued fraction of $G_z / M$, and thus $\bar{A}_y = A_{y_1} \ldots A_{y_h}$ can be found in polytime. □

Notice that (4') is insufficient for $\ell(x_1) + \ldots + \ell(x_m) = \ell(y_1) + \ldots + \ell(y_h)$ (see Property 7), and $\bar{A}_y$ is faced with nonuniqueness because there may possibly exist several convergents of the continued fraction of $G_z / M$ which all satisfy (4').

**Property 4 (Indeterminacy of $\ell(.)$)**: Let $h + m \leq n$. $\forall x_1, \ldots, x_m, y_1, \ldots, y_h \in [1, n]$, and $\|W\| \neq \bar{M}$.

① When $\ell(x_1) + \ldots + \ell(x_m) = \ell(y_1) + \ldots + \ell(y_h)$, and $m \neq h$, there is
$$\ell(x_1) + \|W\| + \ldots + \ell(x_m) + \|W\| \neq \ell(y_1) + \|W\| + \ldots + \ell(y_h) + \|W\| \ (\% \ \bar{M});$$

② when $\ell(x_1) + \ldots + \ell(x_m) \neq \ell(y_1) + \ldots + \ell(y_h)$, there always exist
$$C_{x_1} \equiv A'_{x_1} W'^{\ell(x_1)}, \ldots, C_{x_m} \equiv A'_{x_m} W'^{\ell(x_m)},$$
$$C_{y_1} \equiv A'_{y_1} W'^{\ell(y_1)}, \ldots, C_{y_h} \equiv A'_{y_h} W'^{\ell(y_h)} \ (\% \ M),$$
such that $\ell'(x_1) + \ldots + \ell'(x_m) \equiv \ell'(y_1) + \ldots + \ell'(y_h) \ (\% \ \bar{M})$ with $A'_{y_1} \ldots A'_{y_h} \leq P^h$;

③ when $\ell(x_1) + \ldots + \ell(x_m) \neq \ell(y_1) + \ldots + \ell(y_h)$, probability that $C_{x_1}, \ldots, C_{x_m}, C_{y_1}, \ldots, C_{y_h}$ make (4) with $A'_{y_1} \ldots A'_{y_h} \leq P^h$ hold is roughly $1 / 2^{n-m-h-1}$.

*Proof.*

① It is easy to understand that
$$W^{\ell(x_1)} \equiv W^{\ell(x_1) + \|W\|}, \ldots, W^{\ell(x_m)} \equiv W^{\ell(x_m) + \|W\|} \ (\% \ M),$$
$$W^{\ell(y_1)} \equiv W^{\ell(y_1) + \|W\|}, \ldots, W^{\ell(y_h)} \equiv W^{\ell(y_h) + \|W\|} \ (\% \ M),$$

Due to $\|W\| \neq \bar{M}$, $m\|W\| \neq h\|W\|$, and $\ell(x_1) + \ldots + \ell(x_m) = \ell(y_1) + \ldots + \ell(y_h)$, it follows that
$$\ell(x_1) + \ldots + \ell(x_m) + m\|W\| \neq \ell(y_1) + \ldots + \ell(y_h) + h\|W\| \ (\% \ \bar{M}).$$

② Because $A'_{y_1} \ldots A'_{y_h}$ need be observed, the constraint $A'_{y_1} \ldots A'_{y_h} \leq P^h$ is demanded while because $A'_{x_1}, \ldots, A'_{x_m}$ need not be observed, the constraints $A'_{x_1} \leq P, \ldots, A'_{x_m} \leq P$ are not demanded.

Let $\bar{O}_d$ be an oracle on a discrete logarithm.

Suppose that $W' \in [1, \bar{M}]$ is a generator of $(\mathbb{Z}_M^*, \cdot)$.

Let $\mu = \ell'(y_1) + \ldots + \ell'(y_h)$. In terms of group theories, $\forall A'_{y_1}, \ldots, A'_{y_h} \in [2, P]$ which need not be pairwise coprime, the equation

$$C_{y_1}\ldots C_{y_h} \equiv A'_{y_1}\ldots A'_{y_h} W'^{\mu} \ (\% \ M)$$

in $\mu$ has a solution. $\mu$ may be obtained through $\bar{O}_d$.

$\forall \ \ell'(y_1), \ldots, \ell'(y_{h-1}) \in [1, \bar{M}]$, let $\ell'(y_h) \equiv \mu - (\ell'(y_1) + \ldots + \ell'(y_{h-1})) \ (\% \ \bar{M})$.

Similarly,

$\forall \ \ell'(x_1), \ldots, \ell'(x_{m-1}) \in [1, \bar{M}]$, let $\ell'(x_m) \equiv \mu - (\ell'(x_1) + \ldots + \ell'(x_{m-1})) \ (\% \ \bar{M})$.

Further, from $Cx_1 \equiv A'_{x_1} W'^{\ell(x_1)}, \ldots, Cx_m \equiv A'_{x_m} W'^{\ell(x_m)} \ (\% \ M)$, we can obtain a tuple $(A'_{x_1}, \ldots, A'_{x_m})$, where $A'_{x_1}, \ldots, A'_{x_m} \in (1, M)$, and $\ell'(x_1) + \ldots + \ell'(x_m) \equiv \ell'(y_1) + \ldots + \ell'(y_h) \ (\% \ \bar{M})$.

Thus, Property 4.1 is proven.

③ Let $G_z \equiv Cx_1\ldots Cx_m (Cy_1\ldots Cy_h)^{-1} \ (\% \ M)$. Then in terms of Property 4.1, there is

$$Cx_1\ldots Cx_m (Cy_1\ldots Cy_h)^{-1} \equiv A'_{x_1}\ldots A'_{x_m} W'^{\ell(x_1)+\ldots+\ell(x_m)} (A'_{y_1}\ldots A'_{y_h} W'^{\ell(y_1)+\ldots+\ell(y_h)})^{-1}$$

with $\ell'(x_1) + \ldots + \ell'(x_m) \equiv \ell'(y_1) + \ldots + \ell'(y_h) \ (\% \ \bar{M})$.

Further, there is

$$A'_{x_1}\ldots A'_{x_m} \equiv Cx_1\ldots Cx_m (Cy_1\ldots Cy_h)^{-1} A'_{y_1}\ldots A'_{y_h} \ (\% \ M).$$

The above equation manifests that the values of $W'$ and $(\ell'(y_1) + \ldots + \ell'(y_h)$ or $\ell'(x_1) + \ldots + \ell'(x_m))$ do not influence the value of the product $A'_{x_1}\ldots A'_{x_m}$.

If $A'_{y_1}\ldots A'_{y_h} \in [2^h, \mathcal{P}^h]$ changes, the product $A'_{x_1}\ldots A'_{x_m}$ also changes, where $A'_{y_1}\ldots A'_{y_h}$ is a composite integer. Therefore, $\forall \ x_1, \ldots, x_m, y_1, \ldots, y_h \in [1, n]$, the number of potential values of $A'_{x_1}\ldots A'_{x_m}$ is roughly $(\mathcal{P}^h - 2^h + 1)$.

Let $M = q\mathcal{P}^m (A'_{y_1}\ldots A'_{y_h}) 2^{n-m-h}$, where $q$ is a rational number.

According to (3),

$$G_z / M - L / (A'_{y_1}\ldots A'_{y_h}) = (A'_{x_1}\ldots A'_{x_m}) / (M A'_{y_1}\ldots A'_{y_h})$$
$$= (A'_{x_1}\ldots A'_{x_m}) / (q\mathcal{P}^m 2^{n-m-h} (A'_{y_1}\ldots A'_{y_h})^2).$$

When $A'_{x_1}\ldots A'_{x_m} \leq q\mathcal{P}^m$, there is

$$G_z / M - L / (A'_{y_1}\ldots A'_{y_h}) \leq q\mathcal{P}^m / (q\mathcal{P}^m 2^{n-m-h} (A'_{y_1}\ldots A'_{y_h})^2)$$
$$= 1 / (2^{n-m-h} (A'_{y_1}\ldots A'_{y_h})^2),$$

which satisfies (4).

Assume that the value of $A'_{x_1}\ldots A'_{x_m}$ distributes uniformly on the interval $(1, M)$. If $A'_{y_1}\ldots A'_{y_h}$ is a certain concrete value, the probability that $A'_{x_1}\ldots A'_{x_m}$ makes (4) hold at a specific value of $A'_{y_1}\ldots A'_{y_h}$ is

$$q\mathcal{P}^m / M = q\mathcal{P}^m / (q\mathcal{P}^m (A'_{y_1}\ldots A'_{y_h}) 2^{n-m-h})$$
$$= 1 / ((A'_{y_1}\ldots A'_{y_h}) 2^{n-m-h}).$$

In fact, it is possible that $A'_{y_1}\ldots A'_{y_h}$ take every value in the interval $[2^h, \mathcal{P}^h]$ when $Cx_1, \ldots, Cx_m, Cy_1, \ldots, Cy_h$ are fixed. Thus, the probability that $A'_{x_1}\ldots A'_{x_m}$ makes (4) hold is

$$P_{\forall x_1, \ldots, x_m, y_1, \ldots, y_h \in [1,n]} = (1/(2^{n-m-h}))(1/2^h + 1/(2^h+1) + \ldots + 1/\mathcal{P}^h)$$
$$> (1/2^{n-m-h})(2(\mathcal{P}^h - 2^h + 1)/(\mathcal{P}^h + 2^h))$$
$$= (\mathcal{P}^h - 2^h + 1)/(2^{n-m-h-1}(\mathcal{P}^h + 2^h))$$
$$\approx 1/2^{n-m-h-1}.$$

Obviously, the larger $m + h$ is, the larger the probability is, and the smaller $n$ is, the larger the probability is also. □

**Property 5**: Let $h + m \leq n$. $\forall \ x_1, \ldots, x_m, y_1, \ldots, y_h \in [1, n]$, when $\ell(x_1) + \ldots + \ell(x_m) = \ell(y_1) + \ldots + \ell(y_h)$, the probability that another $\bar{A}_y$ makes (4) with $\bar{A}_y \leq \mathcal{P}^h$ hold is roughly $1/2^{n-m-h-1}$.

*Proof.*

Let

$$G_x \equiv Cx_1\ldots Cx_m \equiv (Ax_1\ldots Ax_m) W^{\ell(x_1)+\ldots+\ell(x_m)} \ (\% \ M),$$
$$G_y \equiv Cy_1\ldots Cy_h \equiv (Ay_1\ldots Ay_h) W^{\ell(y_1)+\ldots+\ell(y_h)} \ (\% \ M).$$

Due to $\ell(x_1) + \ldots + \ell(x_m) = \ell(y_1) + \ldots + \ell(y_h)$, there is

$$G_z \equiv G_x G_y^{-1} \equiv (Ax_1\ldots Ax_m)(Ay_1\ldots Ay_h)^{-1} \equiv (Ax_1\ldots Ax_m)\bar{A}_y^{-1} \ (\% \ M).$$

According to the derivation of (4″), $\bar{A}_y$ will occur in a convergent of the continued fraction of $G_z / M$.

Let $p_1 / q_1, \ldots, p_x / q_x = L / \bar{A}_y, p_{x+1} / q_{x+1}, \ldots, p_t / q_t$ be the convergent sequence of the continued fraction of $G_z / M$, where $t \leq \lceil \lg M \rceil$.

Because of $G_z / M - L / \bar{A}_y < 1/(2^{n-m-h} \bar{A}_y^2)$, it will lead

$$|G_z / M - p_{x+1} / q_{x+1}| < 1/(2^{n-m-h} q_{x+1}^2) \text{ with } q_{x+1} \leq \mathcal{P}^h,$$

$$\ldots\ldots, \text{or}$$

$$|G_z / M - p_t / q_t| < 1/(2^{n-m-h} q_t^2) \text{ with } q_t \leq \mathcal{P}^h$$

to probably hold, and in terms of Property 4.2, the probability is roughly $1/2^{n-m-h-1}$.

Notice that in this case, there is $\ell'(x_1) + \ldots + \ell'(x_m) \equiv \ell'(y_1) + \ldots + \ell'(y_h) \ (\% \ \bar{M})$ with $A'_{y_1}\ldots A'_{y_h} \leq \mathcal{P}^h$, where $\ell'(x_1), \ldots, \ell'(x_m), \ell'(y_1), \ldots, \ell'(y_h)$ satisfy
$$Cx_1 \equiv A'x_1 W'^{\ell'(x_1)}, \ldots, Cx_m \equiv A'x_m W'^{\ell'(x_m)}, Cy_1 \equiv A'y_1 W'^{\ell'(y_1)}, \ldots, Cy_h \equiv A'y_h W'^{\ell'(y_h)} \; (\% \, M).$$
End of the proof. □

Property 5 illuminates that the nonuniqueness of $\bar{A}_y$, namely there may exist the disturbance of $\bar{A}_y$. The smaller $m + h$ is, the less the disturbance is.

### 4.2 Some Conditions Are Only Necessary But Not Sufficient

***Property 6***: (4) is necessary but insufficient for $\ell(x_1) + \ldots + \ell(x_m) = \ell(y_1) + \ldots + \ell(y_h)$ with $x_1, \ldots, x_m, y_1, \ldots, y_h \in [1, n]$, namely for the powers of $W$ and $W^{-1}$ in $G_z$ to counteract each other.

*Proof.* Necessity:

Suppose that $\ell(x_1) + \ldots + \ell(x_m) = \ell(y_1) + \ldots + \ell(y_h)$.

Let $\{C_1, \ldots, C_n\}$ be a public key sequence, and $M$ be a modulus, where $C_i \equiv A_i W^{\ell(i)} \; (\% \, M)$.

Let $G_x \equiv Cx_1 \ldots Cx_m \; (\% \, M)$, $G_y \equiv Cy_1 \ldots Cy_h \; (\% \, M)$, and $G_z \equiv G_x G_y^{-1} \; (\% \, M)$.

Further, $G_z \equiv (Ax_1 \ldots Ax_m)(Ay_1 \ldots Ay_h)^{-1} \; (\% \, M)$.

Denote the product $Ay_1 \ldots Ay_h$ by $\bar{A}_y$. Similar to Section 4.1, we have
$$G_z / M - L / \bar{A}_y < 1 / (2^{n-m-h} \bar{A}_y^2),$$
Namely (4) holds.

Insufficiency:

Suppose that (4) holds.

The contrapositive of the proposition that if (4) holds, $\ell(x_1) + \ldots + \ell(x_m) = \ell(y_1) + \ldots + \ell(y_h)$ holds is that if $\ell(x_1) + \ldots + \ell(x_m) \neq \ell(y_1) + \ldots + \ell(y_h)$, (4) does not hold.

Hence, we need to prove that when $\ell(x_1) + \ldots + \ell(x_m) \neq \ell(y_1) + \ldots + \ell(y_h)$, (4) still holds.

In terms of Property 4.2, when $\ell(x_1) + \ldots + \ell(x_m) \neq \ell(y_1) + \ldots + \ell(y_h)$, the (4) holds with the probability $1/2^{n-m-h-1}$, which reminds us that when $\{C_1, \ldots, C_n\}$ is generated, some subsequences in the forms $\{Cx_1, \ldots, Cx_m\}$ and $\{Cy_1, \ldots, Cy_h\}$ which are verified to satisfy (4) with $\ell(x_1) + \ldots + \ell(x_m) \neq \ell(y_1) + \ldots + \ell(y_h)$ can always be found beforehand through adjusting the values of $W$ and some elements in $\{A_1, A_2, \ldots, A_i\}$ or $\{\ell(1), \ell(2), \ldots, \ell(n)\}$.

Hence, the (4) is not sufficient for $\ell(x_1) + \ldots + \ell(x_m) = \ell(y_1) + \ldots + \ell(y_h)$. □

***Property 7***: (4′) is necessary but not sufficient for $\ell(x_1) + \ldots + \ell(x_m) = \ell(y_1) + \ldots + \ell(y_h)$ with $x_1, \ldots, x_m, y_1, \ldots, y_h \in [1, n]$, for the powers of $W$ and $W^{-1}$ in $G_z$ to counteract each other.

*Proof.*

Because (4′) is derived from (4), and Property 6 holds, naturally Property 7 also holds. □

***Property 8***: Let $m = 2$ and $h = 1$. $\forall \, x_1, x_2, y_1 \in [1, n]$, when $\ell(x_1) + \ell(x_2) \neq \ell(y_1)$,

① there always exist
$$Cx_1 \equiv A'x_1 W'^{\ell'(x_1)}, \; Cx_2 \equiv A'x_2 W'^{\ell'(x_2)}, \; Cy_1 \equiv A'y_1 W'^{\ell'(y_1)} \; (\% \, M),$$
such that $\ell'(x_1) + \ell'(x_2) \equiv \ell'(y_1) \; (\% \, \bar{M})$ with $A'y_1 \leq \mathcal{P}$;

② $Cx_1, Cx_2, Cy_1$ make (4″) with $A'y_1 \leq \mathcal{P}$ hold in all probability.

*Proof.*

① It is similar to the proving process of Property 4.1.

② Let
$$G_z \equiv Cx_1 Cx_2 Cy_1^{-1} \equiv A'x_1 A'x_2 W'^{\ell'(x_1) + \ell'(x_2)} (A'y_1 W'^{\ell'(y_1)})^{-1} \; (\% \, M)$$
with $\ell'(x_1) + \ell'(x_2) \equiv \ell'(y_1) \; (\% \, \bar{M})$.

Further, there is $A'x_1 A'x_2 \equiv Cx_1 Cx_2 Cy_1^{-1} A'y_1 \; (\% M)$.

It is easily seen from the above equations that the values of $W'$ and $\ell'(y_1)$ do not influence the value of $(A'x_1 A'x_2)$.

If $A'y_1 \in [2, \mathcal{P}]$ changes, $A'x_1 A'x_2$ also changes. Thus, $\forall \, x_1, x_2, y_1 \in [1, n]$, the number of potential values of $A'x_1 A'x_2$ is $\mathcal{P} - 1$.

Let $M = 2q\mathcal{P}^2 A'y_1$, where $q$ is a rational number.

According to (3),
$$G_z / M - L / A'y_1 = A'x_1 A'x_2 / (M A'y_1) = A'x_1 A'x_2 / (2q\mathcal{P}^2 A'y_1^2).$$

When $A'x_1 A'x_2 \leq q\mathcal{P}^2$, there is
$$G_z / M - L / A'y_1 \leq q\mathcal{P}^2 / (2q\mathcal{P}^2 A'y_1^2) = 1 / (2 A'y_1^2),$$
which satisfies (4″).

Assume that the value of $A'x_1 A'x_2$ distributes uniformly on $(1, M)$. Then, the probability that $A'x_1 A'x_2$ makes (4″) hold is

$$P_{\forall x_1, x_2, y_1 \in [1, n]} = (q\mathcal{P}^2 / (2q\mathcal{P}^2))(1/2 + \ldots + 1/\mathcal{P})$$
$$\geq (1/2)(2(\mathcal{P} - 1)/(\mathcal{P} + 2)) = 1 - 3/(\mathcal{P} + 2).$$

Apparently, $P_{\forall x_1, x_2, y_1 \in [1, n]}$ is very large, and especially when $\mathcal{P}$ is pretty large, it is close to 1. □

According to Property 8.2, for a certain $C_{y_1}$ and $\forall\, C_{x_1}, C_{x_2} \in \{C_1, \ldots, C_n\}$, attack by (4″) will produce roughly $n^2 / 2$ possible values of $A_{y_1}$, including the repeated, while attack by (4) may filter out most of the disturbing data of $A_{y_1}$. Because every $A_{y_1} \leq P < n^2 / 2$ in REESSE1, the number of potential values of $A_{y_1}$ is at most $P$ in terms of the pigeonhole principle, which indicates the running time of discriminating the original coprime sequence from the values of $A_1, \ldots$, the values of $A_n$ is $O(P^n)$.

**Property 9**: (4″) is necessary but not sufficient for $\ell(x_1) + \ell(x_2) = \ell(y_1)$ with $x_1, x_2, y_1 \in [1, n]$, namely for the powers of $W$ and $W^{-1}$ in $G_z$ to counteract each other.

*Proof.*

Because (4″) is derived from (4), and Property 6 holds, naturally Property 9 also holds. □

It should be noted that Property 2, 3, …, 9 do not depend on the selection of codomain of the lever function $\ell(.)$, namely regardless of selecting the old $\Omega$ or the new $\Omega_\pm$, Property 2, 3, …, 9 still hold.

### 4.3 Two Discrepant Cases

The cases of $h = 1$ and $h \neq 1$ need to be treated distinguishingly.

### 4.3.1 Case of $h = 1$: Verification by Examples

The $h = 1$ means that $\bar{A}_y = A_{y_1}$. If $\bar{A}_y$ is determined, a certain $A_{y_1}$ might be exposed directly. A single $A_{y_1}$ may be either prime or composite, and thus "whether $A_{y_1}$ is a prime" may not be regarded as the criterion of the powers of $W$ and $W^{-1}$ counteracting each other.

If take $m = 2$ and $h = 1$, in terms of Property 4.2, the probability $P_{\forall x_1, x_2, y_1 \in [1, n]}$ that $A'_{x_1} A'_{x_2}$ makes (4) hold is roughly $1/2^{n-4}$, and the number of rationals formed as $G_z / M$ which lead (4) to hold is roughly $n^3 / 2^{n-4}$ when the interval $[1, n]$ is traversed by $x_1, x_2, y_1$ separately. Notice that $P_{\forall x_1, x_2, y_1 \in [1, n]}$ is with respect to (4), but not with respect to (4′) or (4″).

Notice that due to $\Omega_\pm = \{\pm 5, \pm 6, \ldots, \pm(n + 4)\}$, the value of $\ell(x_1) + \ell(x_2) = (-5) + 6 = 1$ for example does not necessarily occur in $\Omega_\pm$.

In what follows, we validate Property 6 and 8 with two examples when $m = 2$ and $h = 1$. Especially assume that $\Omega_\pm = \{5, 6, \ldots, n + 4\}$ to a turn is selected.

*Example 1*:

It will illustrate the ineffectuality of continued fraction attack by (4).

Assume that the bit-length of a plaintext block is $n = 6$.

Let $\{A_i\} = \{11, 10, 3, 7, 17, 13\}$, and $\Omega_\pm = \{5, 6, 7, 8, 9, 10\}$.

Find $M = 510931 > 11 \times 10 \times 3 \times 7 \times 17 \times 13$.

Stochastically pick $W = 17797$, and
$$\ell(1) = 9,\ \ell(2) = 6,\ \ell(3) = 10,\ \ell(4) = 5,\ \ell(5) = 7,\ \ell(6) = 8.$$
From $C_i \equiv A_i W^{\ell(i)}\ (\%\ M)$, we obtain $\{C_i\} = \{113101, 79182, 175066, 433093, 501150, 389033\}$.

Stochastically pick $x_1 = 2$, $x_2 = 6$, and $y_1 = 5$. Notice that there is $\ell(5) \neq \ell(2) + \ell(6)$.

Compute
$$G_z \equiv C_2 C_6 C_5^{-1} \equiv 79182 \times 389033 \times 434038 \equiv 342114\ (\%\ 510931).$$

Presuppose that the power of $W$ in $C_2 C_6$ is just counteracted by the power of $W^{-1}$ in $C_5^{-1}$, and then
$$342114 \equiv A_2 A_6 A_5^{-1}\ (\%\ 510931).$$

According to (3),
$$342114 / 510931 - L / A_5 = A_2 A_6 / (510931\, A_5).$$

It follows that the continued fraction expansion of $342114 / 510931$ equals
$$1/(1 + 1/(2 + 1/(37 + 1/(1 + 1/(2 + \ldots + 1/4))))),$$
where the denominators $1 = a_1,\ 2 = a_2,\ 37 = a_3,\ \ldots$ .

Heuristically let
$$L / A_5 = 1/(1 + 1/2) = 2/3,$$
which indicates it is probable that $A_5 = 3$. Further,
$$342114 / 510931 - 2/3 = 0.002922769 < 1/(2^3 \times 3^2) = 0.013888889,$$
which satisfies (4). Then $A_5 = 3$ is deduced, which is in direct contradiction to factual $A_5 = 17$, so it is impossible that (4) may serve as a sufficient condition.

Meantime, in Example 1, we observe $a_2 = 2$ and $a_3 = 37$, and the increase from $a_2$ to $a_3$ should be sharp. However, even though the case is this, continued fraction attack by (4) fails.

*Example 2*:

It will illustrate the ineffectuality of continued fraction attack by a discriminant relevant to (4″).

The following algorithm which is evolved from the analysis task in [12] describes continued fraction attack on a simplified REESSE1+ private key. The attack rests on the discriminant

$$q_s \Delta < q_{s+1} \text{ and } q_s < A_{\max}, \tag{5}$$

where $q_s$, $q_{s+1}$, $\Delta$, and $A_{\max}$ are referred to the following algorithm for their meanings.

In terms of [12], (5) is derived from (4″). Seemingly, (5) is stricter than (4″), and intentionally used uniquely to determine the value of $A_{y_1}$.

INPUT: a public key ($\{C_1, \ldots, C_n\}$, $M$).

S1: Generate the first $2n$ primes $p_1, \ldots, p_{2n}$ of the natural set.

S2: Set $\Delta \leftarrow (M / (2\prod_{i=n-2}^{u} p_i))^{1/2}$, $A_{\max} \leftarrow M / \prod_{i=1}^{n-1} p_i$,
where $u$ meets $\prod_{i=1}^{u} p_i < M \leq \prod_{i=1}^{u+1} p_i$.

S3: For ($x_1 = 1$, $x_1 \leq n$, $x_1$++)
   For ($x_2 = 1$, $x_2 \leq n$, $x_2$++)
   For ($y_1 = 1$, $y_1 \leq n$, $y_1$++)
      S3.1: Compute $G_z \leftarrow C_{x_1} C_{x_2} C_{y_1}^{-1} \% M$;
      S3.2: Get convergent sequence $\{r_0/q_0, r_1/q_1, \ldots, r_t/q_t\}$ of continued fraction of $G_z/M$;
      S3.3: Get denominator sequence $\{q_1, q_2, \ldots, q_t\}$ from the convergent sequence;
      S3.4: For ($s = 1$, $s \leq t$, $s$++)
         If ($q_s \Delta < q_{s+1}$) and ($q_s < A_{\max}$) then
            Let $A_{y_1} \leftarrow q_s$; and output $\langle A_{y_1}, (x_1, x_2, y_1)\rangle$.

S4: Return.

OUTPUT: All tuples $\langle A_{y_1}, (x_1, x_2, y_1)\rangle$.

Notice that a statement $z$++ denotes $z \leftarrow z + 1$, where $z$ is any arbitrary variable.

However, Algorithm 4.3.1 is ineffectual in practice. Please see the following example.

Assume that the bit-length of a plaintext block is $n = 10$.

Let $\{A_i\} = \{437, 221, 77, 43, 37, 29, 41, 31, 15, 2\}$, and $\Omega_\pm = \{5, 6, 7, 8, 9, 10, 11, 12, 13, 14\}$.

Find $M = 130827613316700077 > \prod_{i=1}^{n} A_i = 130827613316700030$.

Randomly select $W = 944516391$, and

$\ell(1) = 11$, $\ell(2) = 14$, $\ell(3) = 13$, $\ell(4) = 8$, $\ell(5) = 10$, $\ell(6) = 5$, $\ell(7) = 9$, $\ell(8) = 7$, $\ell(9) = 12$, $\ell(10) = 6$.

By $C_i \equiv A_i W^{\ell(i)}$ (% $M$), obtain

$\{C_1, \ldots, C_{10}\} = \{3534250731208421, 12235924019299910, 8726060645493642, 10110020851673707, 2328792308267710, 8425476748983036, 6187583147203887, 10200412235916586, 9359330740489342, 5977236088006743\}$.

On input the public key $\langle\{C_i\}, M\rangle$, the program by Algorithm 4.3.1 will evaluate $\Delta = 506$, $A_{\max} = 58642670$, and output $A_{y_1}$ and $(x_1, x_2, y_1)$. Construct Table 1 with entries $\langle A_{y_1}, (x_1, x_2, y_1)\rangle$. On Table 1, the number of triples $(x_1, x_2, y_1)$ is greater than 100.

**Table 1**. $A_{y_1}$ and the Triple $(x_1, x_2, y_1)$

| $A_{y_1}$ | Triple $(x_1, x_2, y_1)$ | $A_{y_1}$ | Triple $(x_1, x_2, y_1)$ |
|---|---|---|---|
| $A_1 = 187125$ | (1, 1, 1) | $A_5 = 187125$ | (5, 1, 5), (1, 5, 5) |
| $A_1 = 121089$ | (2, 1, 1), (1, 2, 1) | $A_5 = 630269$ | (6, 1, 5), (1, 6, 5) |
| $A_1 = 77$ | (5, 3, 1), (3, 5, 1) | $A_5 = 121089$ | (5, 2, 5), (2, 5, 5) |
| $A_1 = 23$ | (8, 6, 1), (6, 8, 1), (10, 10, 1) | $A_5 = 41$ | (8, 2, 5), (2, 8, 5) |
| $A_1 = 437$ | (10, 6, 1), (6, 10, 1) | $A_5 = 97$ | (4, 3, 5), (3, 4, 5) |
| $A_2 = 1251$ | (1, 1, 2) | $A_5 = 37$ | (6, 6, 5), (10, 6, 5), (6, 10, 5) |
| $A_2 = 187125$ | (2, 1, 2), (1, 2, 2) | $A_6 = 187125$ | (6, 1, 6), (1, 6, 6) |
| $A_2 = 121089$ | (2, 2, 2) | $A_6 = 121089$ | (6, 2, 6), (2, 6, 6) |
| $A_2 = 17$ | (8,4,2), (6,5,2), (5,6,2), (10,7,2), (4,8,2), (7,10,2) | $A_7 = 187125$ | (7, 1, 7), (1, 7, 7) |
| $A_2 = 221$ | (10, 4, 2), (7, 6, 2), (6, 7, 2), (8, 8, 2), (4, 10, 2) | $A_7 = 121089$ | (7, 2, 7), (2, 7, 7) |
| $A_2 = 77$ | (9, 8, 2), (8, 9, 2) | $A_7 = 3$ | (9, 3, 7), (3, 9, 7) |
| $A_2 = 4204$ | (10, 10, 2) | $A_8 = 187125$ | (8, 1, 8), (1, 8, 8) |
| $A_3 = 187125$ | (3, 1, 3), (1, 3, 3) | $A_8 = 34945619$ | (6, 2, 8), (2, 6, 8) |
| $A_3 = 12$ | (7, 1, 3), (1, 7, 3) | $A_8 = 121089$ | (8, 2, 8), (2, 8, 8) |
| $A_3 = 121089$ | (3, 2, 3), (2, 3, 3) | $A_9 = 187125$ | (9, 1, 9), (1, 9, 9) |
| $A_3 = 77$ | (6, 4, 3), (4, 6, 3), (10, 8, 3), (8, 10, 3) | $A_9 = 121089$ | (9, 2, 9), (2, 9, 9) |
| $A_3 = 11$ | (10, 4, 3), (7, 6, 3), (6, 7, 3), (8, 8, 3), (4, 10, 3) | $A_9 = 5$ | (6, 4, 9), (4, 6, 9), (10, 8, 9), (8, 10, 9) |
| $A_3 = 2113$ | (8, 7, 3), (7, 8, 3) | $A_9 = 15$ | (8, 6, 9), (6, 8, 9), (10, 10, 9) |
| $A_3 = 769$ | (9, 8, 3), (8, 9, 3) | $A_{10} = 259970$ | (4, 1, 10), (1, 4, 10) |
| $A_4 = 187125$ | (4, 1, 4), (1, 4, 4) | $A_{10} = 187125$ | (10, 1, 10), (1, 10, 10) |
| $A_4 = 121089$ | (4, 2, 4), (2, 4, 4) | $A_{10} = 121089$ | (10, 2, 10), (2, 10, 10) |
| $A_4 = 76$ | (10, 6, 4), (6, 10, 4) | $A_{10} = 7629$ | (8, 3, 10), (3, 8, 10) |
| $A_4 = 56$ | (10, 9, 4), (9, 10, 4) | | |

On Table 1, we observe that

$A_{y_1}$ relevant to 5 triples is $A_2 = 221$ or $A_3 = 11$,

$A_{y_1}$ relevant to 4 triples is $A_3 = 77$ or $A_9 = 5$,

$A_{y_1}$ relevant to 3 triples is $A_1 = 23$, $A_5 = 37$, or $A_9 = 15$,

$A_{y_1}$ relevant to 2 triples is $A_1 = 77$, $A_2 = 77$, $A_3 = 12$, $A_4 = 56$, $A_5 = 41$, or $A_7 = 3$ etc,

$A_{y_1}$ relevant to 1 triple is $A_1 = 187125$, $A_2 = 1251$, $A_2 = 121089$, or $A_2 = 4204$.

Among these $A_{y_1}$'s, there exist at least $2^{n-5}$ compatible selections from which some elements of the coprime sequence $\{A_i\}$ can be obtained.

For instance, randomly select compatible $A_{y_1}$'s: $A_3 = 11$, $A_9 = 5$, $A_1 = 23$, $A_5 = 41$, and $A_2 = 1251$, and work out $\ell(y_1)$'s: $\ell(3) = 14$, $\ell(9) = 13$, $\ell(1) = 12$, $\ell(5) = 11$, and $\ell(2) = 10$ according to the rule that the number of the triples $(x_1, x_2, y_1)$ tied to $A_{y_1}$ equals $(\ell(y_1) - 9)$ when $\ell(y_1) \geq 10$ [12].

Obviously, such $A_1, A_2, A_3, A_5, A_9$ are not original elements, which indicates (5) derived from (4″) is essentially insufficient even if a concrete $\Omega_\pm = \{5, \ldots, n+4\}$ is selected and known.

### 4.3.2 Case of $h \neq 1$

The $h \neq 1$ means $\bar{A}_y = A_{y_1} \ldots A_{y_h}$. It is well known that any composite $\bar{A}_y \neq p^k$ ($p$ is a prime) can be factorized into some prime multiplicative factors, and many coprime sequences of the same length can be obtained from a prime factor set.

For instance, let $h = 3$ and $\bar{A}_y = 210$ with the prime factor set $\{2, 3, 5, 7\}$. We can obtain the coprime sequences $\{5, 6, 7\}$, $\{6, 5, 7\}$, $\{3, 7, 10\}$, $\{10, 3, 7\}$, $\{2, 15, 7\}$, $\{3, 2, 35\}$, etc. Which is the original?

Property 4 makes it clear that due to the indeterminacy of $\ell(.)$, no matter whether the power of $W$ and $W^{-1}$ counteract each other or not, in some cases, one or several values of $\bar{A}_y$ which may be written as the product of $h$ coprime integers, and satisfy (4) can be found out from the convergents of the continued fraction of $G_z/M$ when the interval $[1, n]$ is traversed respectively by $x_1, \ldots, x_m, y_1, \ldots, y_h$. Thus, "whether $\bar{A}_y$ can be written as the product of $h$ coprime integers" may not be regarded as a criterion for verifying that the powers of $W$ and $W^{-1}$ counteract each other.

Moreover, even if the $k$ values $v_1, \ldots, v_k$ of the product $A_{y_1} A_{y_2} \ldots A_{y_h}$ are obtained, where $y_1$ is fixed, and $y_2, \ldots, y_h$ are varied, $\gcd(v_1, \ldots, v_k)$ can not be judged to be $A_{y_1}$ in terms of the definition of a coprime sequence.

If take $m = 2$ and $h = 2$, in terms of Property 4.2 and $P_{\forall x_1, x_2, y_1, y_2 \in [1, n]}$, the number of rationals formed as $G_z/M$ which leads (4) to hold is roughly $n^4 / 2^{n-5}$ when the interval $[1, n]$ is traversed by $x_1, x_2, y_1, y_2$ respectively. What is most pivotal is that the value of $\ell(x_1) + \ell(x_2)$ or $\ell(y_1) + \ell(y_2)$ $\forall x_1, x_2, y_1, y_2 \in [1, n]$ does not necessarily occur in a concrete $\Omega_\pm$.

### 4.4   Time Complexities of Two Attack Tasks

Continued fraction attack and W-parameter intersection attack on $C_i \equiv A_i W^{\ell(i)} (\% M)$ are the two most efficient algorithms at present.

### 4.4.1 Time Complexity of Continued Fraction Attack

It can be seen from section 4.1 that continued fraction attack is based on the assumption that $\ell(x_1) + \ldots + \ell(x_m) = \ell(y_1) + \ldots + \ell(y_h)$. For convenience, usually let $m = 2$ and $h = 1$.

If $\Omega_\pm$ is determined as $\{5, \ldots, n+4\}$, continued fraction attack by (4), (4′), (4″) or (5) contains five steps dominantly.

Note that it is known from Example 2 that $\Omega_\pm = \{5, \ldots, n + 4\}$ does not mean that continued fraction attack described with the following algorithm will succeed.

INPUT: a public key $\langle\{C_1, \ldots, C_n\}, M\rangle$; the set $\Omega_\pm = \{5, \ldots, n + 4\}$.
S1: Structure Table 2 according to $\Omega_\pm$.
S2: Get entries $\langle A_{y_1}, (x_1, x_2, y_1)\rangle$ by calling Algorithm 4.3.1.
S3: Structure Table 1 with entries $\langle A_{y_1}, (x_1, x_2, y_1)\rangle$.
S4: Find coprime $A_{y_1}$ according to Table 1 and Table 2.
S5: Find pairwise different $\ell(y_1)$ according to $A_{y_1}$ and Table 2.
OUTPUT: coprime values of $A_{y_1}$; pairwise different values of $\ell(y_1)$.

**Table 2**. Number of $\ell(x_1) + \ell(x_2) = \ell(y_1)$ over $\Omega_\pm = \{5, \ldots, n + 4\}$

| | $\ell(y_1)$ | 10 | 11 | …… | $n + 4$ |
|---|---|---|---|---|---|
| | $\ell(x_1) + \ell(x_2)$ | 5 + 5 | 5 + 6, 6 + 5 | …… | $5 + (n - 1), \ldots, (n - 1) + 5$ |
| Number of $\ell(y_1) = \ell(x_1) + \ell(x_2)$ | | 1 | 2 | …… | $n - 5$ |

At S4, finding coprime values of $A_{y_1}$ will probably take $O(2^{n-5})$ running time.

At S1, when $\Omega_\pm$ is indeterminate (in fact $\Omega_\pm$ is one of $2^n$ potential sets), an adversary must firstly determine all the elements of $\Omega_\pm$, which will take $O(2^n)$ running time.

### 4.4.2 Time Complexity of W-parameter Intersection Attack

Due to $C_i \equiv A_i W^{\ell(i)} \,(\%\, M)$ with $A_i \in \Lambda = \{2, \ldots, \mathcal{P}\}$ and $\ell(i) \in \Omega_\pm = \{\pm 5, \ldots, \pm(n + 4)\}$ for $i = 1, \ldots, n$, and elements in the sets $\Lambda$ and $\Omega_\pm$ being small, an adversary may attempt the following attack algorithm with indeterminacy.

INPUT: a public key $\langle\{C_1, \ldots, C_n\}, M\rangle$; the set $\Lambda$.
S1: For $i = 1, \ldots, n$ do
  While $\ell(i)$ traverses $\{-5, \ldots, -(n + 4), 5, \ldots, n + 4\}$ do
    While $A_i$ traverses $\Lambda$ do
      S1.1: Compute $W$ such that $W^{\ell(i)} \equiv C_i A_i^{-1} \,(\%\, M)$;
      S1.2: Insert every possible triple $(W, A_i, \ell(i))$ into the set $\overline{V}_i$.
S2: Seek the intersection $\overline{V} = \overline{V}_1 \cap \ldots \cap \overline{V}_n$ on $W$.
S3: If $W$ is unique in $\overline{V}$ and related $(A_i, \ell(i))$ unique in every $\overline{V}_i$ then
    S3.1: Extract a private key $\langle\{A_i\}, \{\ell(i)\}, W\rangle$.
  else (namely $W$ nonunique in $\overline{V}$, or $(A_i, \ell(i))$ nonunique in some $\overline{V}_i$)
    S3.2: Check whether every possible $\{A_1, \ldots, A_n\}$ is a coprime sequence;
    S3.3: Check whether every possible $\{\ell(1), \ldots, \ell(n)\}$ is a lever function.
S4: Arrange valid private keys $\langle\{A_i\}, \{\ell(i)\}, W\rangle$.
OUTPUT: A list of valid private keys $\langle\{A_i\}, \{\ell(i)\}, W\rangle$.

When the number of valid private keys is larger than 1, every valid private key needs to be verified in order to find the original private key.

Notice that (1) at S1.1, we may compute $W$ by the Moldovyan root finding method the time complexity of which is $O((\max(\ell(i)))^{1/2}\lceil \lg M \rceil) \approx O(n^{1/2}\lceil \lg M \rceil)$ [13]; (2) in the identical set $\overline{V}_i$, to the identical value of $W$, there may exist different related values of $(A_i, \ell(i))$.

The size of every $\overline{V}_i$ is about $O(|\Lambda|\,|\Omega_\pm|/2) \approx O(\mathcal{P}n^2)$ due to $q^2 \mid \overline{M} \,\forall\, q$ (a prime) $\in |\Omega_\pm|$.

At S2, seeking the intersection $\overline{V}$ will take $O(\mathcal{P}n^3)$ running time which is polynomial in $n$.

At S3, seeking a coprime sequence will take $O(n)$ running time in the best case with pretty low probability due to $q^2 \mid \overline{M} \,\forall\, q$ (a prime) $\in |\Omega_\pm|$, but it will take $O(2^n)$ running time in a worse case.

Thus, the adversary cannot extract a simplified REESSE1+ private key in determinate polytime.

## 5  Relation between a Lever Function and a Random Oracle

### 5.1  What Is a Random Oracle

An oracle is a mathematical abstraction, a theoretical black box, or a subroutine of which the running time may not be considered [11][14]. In particular, in cryptography, an oracle may be treated as a subcomponent of an adversary, and lives its own life independent of the adversary. Usually, the adversary interacts with the oracle but cannot control its behavior.

A random oracle is an oracle which answers to every query with a completely random value chosen uniformly from its output domain, except that for any specific query, it outputs the same value every time it receives that query if it is supposed to simulate a deterministic function [14].

In fact, it draws attention that certain artificial signature and encryption schemes are proven secure in the random oracle model, but are trivially insecure when any real hash function such as MD5 or SHA-1 is substituted for the random oracle [15][16]. Nevertheless, for any more natural protocol, a proof of security in the random oracle model gives very strong evidence that an attacker have to discover some unknown and undesirable property of the hash function used in the protocol.

A function or algorithm is regarded random if its output depends not only on the input but also on some random ingredients, namely if its output is not uniquely determined by the input. Hence, to a function or algorithm, randomness contains indeterminacy.

### 5.2 Design of a Random Oracle

Correspondingly, the indeterminacy of the new $\ell(i)$ may be expounded in terms of a random oracle.

Suppose that $\bar{O}_d(y, g)$ is an oracle on solving $y \equiv g^x$ (% $M$) for $x$, and $\bar{O}_\ell$ is an oracle on solving $C_i \equiv A_i W^{\ell(i)}$ (% $M$) for $\ell(i)$, where $M$ is a prime, and the value of $i$ is from 1 to $n$.

Let $D$ be a database which stores records $\langle \{C_1, \ldots, C_n\}, M, \{\ell(1), \ldots, \ell(n)\} \rangle$ computed already.

Additionally, if the arrangement position of some $C_i$ is changed, then $\{C_1, \ldots, C_n\}$ is regarded as a distinct sequence.

The structure of $\bar{O}_\ell$ is as following:

INPUT: a public key $\langle \{C_1, \ldots, C_n\}, M \rangle$.
S1: If find $\langle \{C_1, \ldots, C_n\}, M \rangle$ in $D$
    then retrieve $\{\ell(1), \ldots, \ell(n)\}$, goto S6.
S2: Randomly yield coprime $A_1, \ldots, A_n$
    with $A_i \leq P$ and $\prod_{i=1}^{n} A_i < M$.
S3: Randomly pick a generator $W \in \mathbb{Z}_M^*$.
S4: Evaluate $\ell(i)$ by calling $\bar{O}_d(C_i A_i^{-1}, W)$ for $i = 1, \ldots, n$.
S5: Store $\langle \{C_1, \ldots, C_n\}, M, \{\ell(1), \ldots, \ell(n)\} \rangle$ to $D$.
S6: Return $\{\ell(1), \ldots, \ell(n)\}$, and end.
OUTPUT: a sequence $\{\ell(1), \ldots, \ell(n)\}$.
Of course, $\{A_i\}$ and $W$ as side results may be outputted.

Obviously, for the same input $\langle \{C_1, \ldots, C_n\}, M \rangle$, the output is the same, and for a different input, a related output is random and unpredictable.

Since $C_i A_i^{-1}$ is pairwise distinct, and $W$ is a generator, the result $\{\ell(1), \ldots, \ell(n)\}$ will be pairwise distinct. Again according to Definition 2, every $\ell(i) \in [1, \bar{M}]$ may be beyond $\Omega_\pm$. Thus, $\{\ell(1), \ldots, \ell(n)\}$ is a lever function although it is not necessarily the original.

This section explains further why the continued fraction attack by (4), (4′), (4″), or (5) and the W-parameter intersection attack is ineffectual on $C_i \equiv A_i W^{\ell(i)}$ (% $M$).

## 6    Conclusion

Indeterminacy is ubiquitous. For example, for $x + y = z$, given $x = -122$ and $y = 611$, computing $z = 489$ is easy. Contrarily, given $z = 489$, seeking the original $x$ and $y$ is intractable due to indeterminacy in $x + y = z$. Indeterminacy in $C_i \equiv A_i W^{\ell(i)}$ (% $M$) is similar, and triggered by the lever function $\ell(.)$.

Inequation (4) is stricter than (4″) although both (4) and (4″) are only necessary but insufficient for $\ell(x_1) + \ell(x_2) = \ell(y_1)$. Property 4 and 8 show that attack by (4) is more effectual than attack by (4″) theoretically. However, Section 4.3 shows that when $\Omega_\pm = \{\pm 5, \ldots, \pm(n + 4)\}$ is indeterminate, continued fraction attack by (4), (4′), (4″), or (5) will take $O(2^n)$ running time, and is practically infeasible.

Section 4.4.2 manifests that the W-parameter intersection attack cannot extract a private key in determinate polytime although it unveils some lowly probabilistic risk.

Resorting to the transform $C_i \equiv A_i W^{\ell(i)}$ (% $M$), we expound theoretically the effect of the lever function with adequate indeterminacy, but in practice, to acquire the redundant security of a private key and to decrease the modulus length of the cryptoscheme, we suggest that the key transform should be strengthened to $C_i \equiv (A_i W^{\ell(i)})^\delta$ (% $M$) with $\delta \in [2, \bar{M}], A_i \in \Lambda = \{2, \ldots, P\}$, and $\ell(i) \in \Omega_\pm = \{\pm 5, \ldots, \pm(n + 4)\}$ for $i = 1, \ldots, n$ [6][17].

## Acknowledgment

The authors would like to thank the Academicians Jiren Cai, Zhongyi Zhou, Changxiang Shen, Zhengyao Wei, Xicheng Lu, Jinpeng Huai, Huaimin Wang, Andrew C. Yao, and Binxing Fang for their important advice and help.

The authors also would like to thank the Professors Jie Wang, Zhiying Wang, Ronald L. Rivest, Moti Yung, Dingzhu Du, Hanliang Xu, Dengguo Feng, Yixian Yang, Yupu Hu, Wenbao Han, Huaimin Wang, Jianfeng Ma, Zhiqiu Huang, Lusheng Chen, Bogang Lin, Yiqi Dai, Lequan Min, Dingyi Pei, Mulan Liu, Huanguo Zhang, Qibin Zhai, Hong Zhu, Renji Tao, Quanyuan Wu, and Zhichang Qi for their important suggestions and corrections.